\documentclass{article}
\usepackage{amsmath}
\usepackage{latexsym}
\begin{document}
\title{The Beginning of the End of the Anthropic Principle}
\author{Gordon L. Kane\\
Department of Physics,\\
Randall Laboratory,\\
University of Michigan,\\
Ann Arbor,\\
Michigan 48109-1120,\\
USA.\\ \\
Malcolm J. Perry\\
Department of Applied Mathematics and Theoretical Physics,\\
University of Cambridge,\\
Silver Street,\\
Cambridge CB3 9EW,\\ England.\\ \\
Anna N. \.Zytkow\\
Institute of Astronomy,\\
University of Cambridge,\\
Madingley Road,\\
Cambridge CB3 0HA,\\
England.}
\maketitle
\begin{abstract}
We argue that if string theory as an approach to the fundamental laws of
nature
is correct, then there is almost no room for anthropic arguments in
cosmology.  The quark and lepton masses and interaction strengths are
determined.
\end{abstract}
\vfill
\eject
\section{A brief outline of the anthropic principle}

It is probably fair to say that the existence of humans is an
indisputable fact, ({\it pace} Sartre). Yet curiously an explanation of
this fact was, for most of history, held to be unnecessary. Until the
time of Copernicus, it was widely believed that humanity was the center
of the Universe, and the Universe was made for it, ideas exemplified by
many creation myths in a wide variety of cultures. At least in western
civilization, such an established view was overturned when Copernicus
demonstrated that the Earth was in orbit around the Sun, and thus put on
an equal footing with the other visible planets. Although there was no
fundamental explanation for his observations, it removed humanity from a
central place in the universe. The Earth became simply one of the six
then known planets.  The reason behind Copernicus' observations was
found by Newton. The laws of universal gravitation and mechanics allowed
for an explanation of the gross structure of the Solar system and
also permitted one of the first {\it anthropic} questions to be
asked. Why is it that the Earth is in an orbit with a mean distance from
the Sun, $r_\oplus \sim 1.50\times 10^{13}$ cm, and a low eccentricity,
$e \sim 0.016$? The advantages for life on Earth are easy to see: these
orbital parameters provide a stable temperate environment in which human life,
as we know it, can exist comfortably. A relatively small change in
$r_\oplus$ would lead to an Earth that is either too cold or too hot,
and a change in $e$ to a situation in which there were violent swings in
temperature between the seasons.

As more has been understood, it has been noticed that, apparently, a
number of features of the universe have to be more or less ``just so'',
or humans would not exist. Specific examples have been offered for over
a century (as reviewed in Barrow and Tipler, \cite{BT:cap}), and include
noting that the Sun has to be very stable, the Earth cannot be too small
(else it could not hold an atmosphere) or too large (as gravity would
effectively crush organisms made of molecules), the universe has to be
large and dark and old for life to exist because at least two and
probably three generations of stars are needed to make the heavy
elements life depends on, etc.

The essential point made in anthropic arguments is that we should not be
surprised that the Earth is where it is, because if it were in a
different orbit, then we would not be here to ask the question. The
question is, by virtue of its self-referential nature, nugatory. That
does not mean the question is not worth asking. It may well be that in
the future, our understanding of how the Solar system was formed would
enable us to argue successfully that terrestrial planets are inevitable
in stellar systems of our type. Equally well, they may be unusual
phenomena.

Carter~ \cite{BC:anth1}, appears to be the first to formalize what is meant
by the anthropic principle. He described three types of scientific
reasoning. One is ``traditional.'' Arguments based on our existence are
regarded as extra-scientific. The laws of nature are used to make
predictions in a deductive way. There is some degree of arbitrariness
involved because it is not exactly clear what the laws of nature are,
what the constants of nature are, and what choices of boundary conditions
or quantum states are to be made. In contrast, reasoning based on the
``weak'' anthropic principle allows us to place restrictions on what
we are going to consider to be realistic. Our existence as observers is
privileged in both space and time by virtue of our own existence.
The weak anthropic principle is interesting and unobjectionable. It adds
to our insights, but does not preclude a fully scientific explanation
of any feature of the universe, including the origin of the universe
and the understanding of why the laws of physics are what they are,
with such explanations not depending in any way on knowing whether
humans exist.
What this is saying really is that our existence in a recognizable form
is intimately related to the conditions prevailing now in our part of
the galaxy. ``We are here because we are here,'' would be the sound-bite
associated with this attitude. It has not much predictive power
except that since the Solar system does not seem to be particularly
unusual in any way then it would be reasonable to suppose that life,
and indeed quite possibly some form of civilization,
should also be common at  the
present epoch. (A most un-anthropic conclusion).

The historical starting point, within the context of modern physics,
for anthropic reasoning is to explain  the so-called
large number coincidence, first formalized by Dirac, \cite{D:ln}.
Three large dimensionless quantities, all taking values $\sim 10^{40}$, can
be
found in cosmology. The first is the dimensionless gravitational coupling
referred to the proton mass $m_p,$
\begin{equation}
\frac{\hbar c}{Gm_p^2} \sim 2 \times 10^{38}
\end{equation}
The second is the Hubble time, $T$, referred to the same scale,
 \begin{equation}
\frac{T m_p c^2}{\hbar} \sim 6 \times 10^{41}
\end{equation}
The final quantity is a measure of the mass $M$ of the visible universe
also referred to the same scale
\begin{equation}
\sqrt{\frac{M}{ m_p}} \sim 5 \times 10^{39}
\end{equation}

At first sight, the similarity of these numbers can either be regarded
as a incredible coincidence or a deep fact. But in reality we should
not be surprised. As was first explained by Dicke \cite{RHD:ln},
in a big-bang cosmology these relations are perfectly natural. The age of
the Universe must be in a certain range: not too young as described
earlier,
but also not so old that
stars have largely exhausted their hydrogen fuel.
Dicke showed that the above bounds on the age of the universe were
equivalent to the coincidence of the numerical values  (1)
and (2). The equivalence of (2) and (3) can be interpreted as saying that
we must live in a universe  with a density close to the critical
density. The natural explanation of the second equivalence  is then
the existence of an inflationary epoch.

However, anthropic reasoning can be rather dangerous since it has
a tendency to lead one to draw conclusions that can be theological in
nature. When some phenomenon cannot be understood simply within
a particular system, it is tempting to ascribe its origin as
supernatural, whereas a deeper understanding of physics may allow
a perfectly rational description. The history of physical
science is littered with examples, from medieval times  up to the present.

For some people anthropic reasoning even leads to a ``strong form'' of
the anthropic principle, which argues that our existence places strong
restrictions on the types of theories that can be considered to explain
the universe and the laws of physics, as well as the fundamental
constants of nature. Some would even like to argue that the universe in
some sense had to give rise to humans, or even was designed to do
that. Many physicists are antagonistic to any version of these stronger
arguments. While it is not yet established that the origin of the
universe or the origin of the laws of physics can be understood
scientifically, attempts to answer those questions are finally in the
past decade or so topics of scientific research rather than
speculation. It could happen that such questions do not have scientific
answers, but until the effort is made and fails we will not know.

One approach to partially unifying the laws of physics is, so-called,
``Grand Unification''. In this approach, the weak, electromagnetic,
and strong forces are unified into one simple gauge group of which
the Standard Model gauge group $SU(3)\times SU(2)\times U(1)$ is a subgroup.
In that case it has recently been observed (Hogan~\cite{CH:justso},
Kane~\cite{GLK:what}) that since the
ratios of the coupling strength are fixed,  in making anthropic
calculations one cannot independently change the strong and
electromagnetic forces. (For example, if the strong force strength is
increased a little, the diproton could be bound and cut off nuclear
burning in stars. It should however be noted that even without unification
effects this issue is
more subtle than is usually stated~\cite{AL:book}).
Further, it would be incorrect to argue, as some have done, that
various ``just so'' probabilities should be multiplied together,
since the underlying physics effects are correlated.

Increasing the strength of the electromagnetic
repulsion is required since its ratio to the strong force strength is
fixed by the theory, and that would decrease the diproton binding, so
the net effect would be small. Thus in Grand Unified Theories
a number of anthropic effects disappear.

Recently, Hogan \cite{CH:justso} has argued that some basic microscopic
physics {\it should} be determined anthropically. He suggests that the Grand
Unified theories go about as far as one can hope or expect to go in
relating and explaining the fundamental parameters of a basic theory,
and emphasizes that the sensitivity of the properties of the universe to
a few quantities is very strong. His argument highlights the fact that
in Grand Unified theories there are a number
of independent parameters (particularly quark and lepton masses) that
need to be specified.  He argues that it is important that at least some
of these parameters cannot be determined by the theory, or else we cannot 
understand why the universe is ``just so''.

The context of the above suggestion is that we are living in one of many
universes where these numbers are chosen at random.  The reason we see
them as being what they are is that if they were different, even very
slightly in some cases, then we could not possibly exist. It could be
that these many universes are real and emerge as baby universes in some
meta-universe as yet unobserved, or simply as a statistical ensemble of
distinct universes. In either case, one is faced with the real
difficulty of accounting for why the parameters of the Grand Unified
theory are what they are. It turns out that although at first sight there
appear to a myriad of what might be called anthropic coincidences, 
only four of the parameters appear to be
particularly critical,~\cite{RC:anth}.
They are $m_e, m_u, m_d$ and $g$, the mass of
the electron, up and down quarks respectively, and the Grand Unified
coupling constant (which determines the strengths of the strong,
electromagnetic, and weak forces).  Hogan makes the interesting claim
that if these parameters were determined by the theory, it would be very
hard to understand why the universe is ``just so.''

Today, there is a more ambitious approach to understanding and
unifying the laws of physics, loosely called ``string theory.''
We would like to argue precisely the opposite point of view
to the one presented by Hogan, based on
what is known about string theory, or perhaps more precisely
its non-perturbative progenitor, M-theory.

By string theory we mean the effective 10
dimensional theory that incorporates gravity and quantum theory and the
particles and forces of the Standard Model of particle physics.  Whether
string theory really describes our world is not yet known.  This is
certainly the first time in history when we have a theory to study which
could unify and explain all of nature, conceivably providing an
inevitable primary theory.  Testing string theory ideas may be difficult
but is not in any sense excluded -- one does not need to be present at
the big bang, nor does one need to do experiments at the Planck
scale to test them.

Our goals in writing this paper are first to stress that in string
theories all of the parameters of the theory -- in particular all
quark and lepton masses, and all coupling strengths -- are calculable,
so there are no parameters left to allow anthropic arguments of the
normal kind, or to allow the kind of freedom that Hogan has argued for.
Second, we want to discuss in what ways, if any, there is room to
account for why the universe is ``just so''.

\section{The String Theory picture of low energy physics}

In non-gravitational physics, the role of spacetime is quite clear. It
provides an arena, Minkowski spacetime, in which calculations can be
carried out. In classical gravitational physics, spacetime continues to
exist, but the backgrounds are in general more exotic, representing
diverse situations such as black holes or cosmological models.  It is
easy to graft onto this edifice the content of Grand Unified
theories. However, the philosophy of string unification is to unify all
the forces, including gravitation. At some level, one can successfully
omit gravitation because the natural scale associated with it is $\sim
10^{19} GeV$ whereas the other forces become unified at scales
noticeably less, around $10^{16} GeV$. But, if we are to explore
energies beyond the unification scale, because we want a general theory,
then gravity will become more important and must be included in the
overall picture. To do this, one requires a theory of quantum
gravity. Treating the gravitational field like a gauge theory leads to
an unrenormalizable theory.  To include gravitation, one has to resort
to a theory of extended objects: strings. One way to think about string
theory is to describe the string as an extended object propagating in a
fixed background spacetime.  The metric, or gravitational field, is just
one of the massless degrees of freedom of the string, and it is possible
to extend this picture to include backgrounds that correspond to any of
the massless degrees of freedom of the string. One ingredient of string
theory is that it is described by a conformally invariant theory on the
string world-sheet. This requirement imposes a strict consistency
condition on the allowed backgrounds in which the string lives. The
backgrounds must have ten spacetime dimensions, and obey the
supergravity equations of motion.  We therefore regard string theory as
a consistent quantum theory of gravity in the sense that the theory of
fluctuating strings (including excitations of the string that correspond
to gravitons) is finite provided that the background obeys the equations
of the supergravity theory that corresponds to particular string theory
under discussion.

Next, string theory needs to make contact with the known structure of
the universe. There are, apparently, four spacetime dimensions.
The six remaining directions of spacetime in string theory need to be
removed by a process usually termed compactification. The endpoint of
this process usually arrives at simple ($N=1$) supergravity
coupled to various matter fields.
As a consequence, a severe restriction applies as to how
the compactification takes place. One assumes that spacetime takes
the form of ${\cal M}^4\times K,$ where ${\cal M}^4$ is some
four-dimensional Lorentzian  manifold, and $K$ is a compact space with
six real dimensions. In order to have unbroken simple ($N=1$) supersymmetry,
there is a severe restriction on the nature of $K$.
$K$ must be a so-called ``Calabi-Yau'' manifold and its
spatial extent must be sufficiently
small that it has no direct observational signature,
which restricts it to
scales of around $10^{-30}$ cm, or roughly the Planck scale~\cite{W:cy}.

This in fact is an example of the weak anthropic principle at work.
There seems no reason why one should compactify down to 
four dimensions. From the string theory perspective, 
there is nothing wrong with having
a spacetime of any number of dimensions less than or equal to ten.
However, one could certainly not have intelligent life in either one or
two spatial dimensions. It is impossible to have a complicated
interconnected set of nerve cells unless the number of space dimensions
is at least three, since otherwise one is forced only to have nearest
neighbor connections. In spatial dimensions greater than three, even if
one could have stars, one could not have stable Newtonian planetary
orbits or stable atoms~\cite{PE:stable}, and it is presumably impossible
for a suitable environment for life to exist. More generally, the
compactified dimensions could vary in size.

The Calabi-Yau space is determined by two distinct types of property.
Firstly we must specify its topology and then  specify the metric on it.
The topology determines at least two important properties of the low-energy
theory, the number of generations and the Yukawa couplings. The metric
can just be one of a family of metrics on the given manifold. The fact
that there appears to be some arbitrariness here is reflected in the
presence of massless scalars, or moduli fields, in the low-energy theory.

If one were just doing field theory, then the above considerations would
really be all that there is to it. However, the geometry of string
theory is rather more interesting. In Riemannian geometry, one cannot
continuously and smoothly deform a metric so as it interpolates between
metrics on two topologically distinct manifolds. If one tries to do
this, one ends up  finding that there is some kind of singularity in
the metric that causes the notion of a manifold to break down. In the
background field approximation to string theory, this will also be
true. However, we also know that there is more to string theory than
that. It appears that one should replace the idea of a classical
background geometry with what is usually called quantum geometry,~\cite{BG:qg}.
Quantum geometry corresponds to those
consistent conformal field theories that define the physics of the
string itself. These conformal field theories are intrinsic to the
string, and have an existence without any concept of spacetime. Thus,
spacetime would be a derived property, rather than being
fundamental. This is closer to the true philosophy of a fundamental
theory, since we should not be trying to draw a distinction between
string physics and the physics of spacetime. A proper discussion of
quantum geometry includes an understanding of non-perturbative effects
in string theory.  However there are two things that we already know for
sure that support the viewpoint of spacetime being a derived property.
The first is the phenomenon of mirror symmetry, ~\cite{BG:elegant}.
It is known that
Calabi-Yau manifolds come in pairs, related by the so-called mirror map,
in which the Kahler and complex structures are interchanged. There is no
obvious connection between the metrics on pairs of mirror manifolds.
However, the conformal field theory associated with the string is in
both cases identical. This indicates that the spacetime description is a
derived one, rather than being fundamental. A second property is that
when non-perturbative phenomena are included, there is no problem from
the string theory point of view in effecting continuous transitions
between Calabi-Yau spaces of {\it different} topology. This shows that
stringy ideas about geometry are really more general than those found in
classical Riemannian geometry. The moduli space of Calabi-Yau manifolds
should thus be regarded as a continuously connected whole, rather than a
series of different ones individually associated with different
topological objects,~\cite{PC:roll}.
Thus, questions about the topology of Calabi-Yau
spaces must be treated on the same footing as questions about the metric
on the spaces. That is, the issue of topology is another aspect of the
the moduli fields.  These considerations are relevant to understanding
the ground state of the universe.

However we reach the final picture of what happens at low energies in
four dimensions, the end point does not contain any massless scalar
fields. That in turn means that when the final $N=1$ supersymmetry is
broken, all of the fields must be associated with effective
potentials. The only exceptions are the genuinely massless fields
associated with unbroken gauge symmetries.  These are the photon for
electromagnetic $U(1)$, the gluons for $SU(3)$ of color, and the
graviton with diffeomorphism invariance.  One could very plausibly
conclude  that  all of the low energy physics must be determined as a
result of this type of compactification plus supersymmetry breaking process.

Thus we have relegated most of the traditionally anthropic quantities to
physics that we know, at least in principle, even if as yet the
calculations are too technically complex to carry out. For example in
the scheme sketched here, it now seems perfectly possible to compute
from first principles quantities like $m_d$ and $m_u$ (see below).  In
fact, it would seem that all of low-energy physics is computable. This
leads us to ask about the small number of possible 
remaining anthropic quantities. For example,
what about the number of the non-compact dimensions
of spacetime. Whilst in string theory, there seems to be no obvious
reason why four should be singled out, in M-theory there is. One's
usual attitude to string theory is that it can be derived from M-theory,
whose low-energy limit is $d=11$ supergravity, by compactifying on a
circle, and then saying that the resultant ten-dimensional spacetime is
the arena for string theory. However, in $d=11$ supergravity theory
there is a four-form field strength that could pick out four dimensions
as being different from the remaining seven if it has a vacuum expectation
value. Thus, a four-seven split seems quite natural. In cosmological
models, the observed universe is described by the
Friedmann-Robertson-Walker models, and they have the property of being
conformally flat. This means that they can be described most simply as a
four-dimensional spacetime of constant curvature, together with a
time-dependent scale factor. Whilst it is not presently understood how
one might realize these ideas in practice, it is beginning to seem
plausible that even something like the dimensionality of the large
directions of spacetime might eventually be understood in M-theory
without recourse to any anthropic arguments at all.  Another observed
fact is that there seems to be a small cosmological constant with a
magnitude similar to the energy density of observed matter in the
Universe. Most string theorists think it is likely, or at least
possible, that a better understanding of string theory will lead to an
understanding of the small size of the cosmological constant, and
possibly also its actual value. It is not inconceivable that a solution to
the dimension problem would come together with a solution to the
cosmological constant problem in M-theory.

\section{The role of parameters in string theory}

As discussed in the previous section, before string theory can be
applied to our actual world several problems have to be solved. They
include compactification to four dimensions, breaking the  full supersymmetry
of the theory that is a hidden symmetry in our world, and finding the
correct ground state (vacuum) of the theory. 
These problems are logically independent,
though it could happen that one insight solves all of them. We assume
here that ongoing research will find solutions to these problems, and
consider the implications for our view of the world and for anthropic
ideas.

Assuming (as described above) that the theory is successfully formulated
as a 4D effective theory near the Planck scale, we describe
qualitatively how force strength and masses are viewed in
string theory.

In the string theory one can think of the theory as having only a
gravitational force in 10D.  The other forces arise in a way analogous
to what happens in the old Kaluza-Klein theory, where one has a 5D world
with only a 5D gravitational force, which splits into a 4D gravitational
force plus electromagnetism when one dimension is compactified.  Thus in
the string theory all of the force strength ratios are fixed by the
structure of the theory.  In string theory the coupling strengths,
including that of gravity,  are
viewed as vacuum expectation values of scalar fields, so the overall
force strengths are calculable too (though how to evaluate those vacuum
expectation values is
not yet known).  If the string approach is correct there is no room for
any coupling strength to vary anthropically.

The situation is similar for masses, though more subtle.  Physical
masses are written as Yukawa couplings (determined by the
compactification) times the electroweak Higgs field vacuum expectation
value.  The quarks and leptons are massless at high temperature scales,
until the universe cools through the electroweak phase transition (at
100 GeV), at which point the quarks and leptons acquire mass.  At higher
scales one speaks of the Yukawa couplings.  The string theory determines
a function called the superpotential, and the coefficients of terms in
the superpotential are the Yukawa couplings.  In general the theory
determines the Yukawa couplings and thus the masses. The way in which
this comes about is directly from the topology of the Calabi-Yau space itself.
Given its topological nature, the Yukawa couplings are determined. 
Thus, there is really not much freedom left at this level.

However, a subtlety may arise.  At the most basic level of the
theory it could happen that some Yukawa couplings were of the same order
as the gauge couplings, giving the masses of the heavier particles such as
the top quark, but some Yukawa couplings corresponding to lighter
particles were zero.  The lighter masses only arise when the full
symmetry group of the theory is broken, perhaps when supersymmetry is
broken or when the compactification occurs.  Then calculating the lighter
masses is technically more challenging.  Nevertheless, it is expected that
in
principle, and eventually in practice, all of the masses are calculable,
including the up and down quark masses, and the electron mass.  There is
not any room for anthropic variation of the masses in a string theory.

\section{What is left that could be anthropic?}

String theory can be any one of the five consistent perturbative
superstring theories that contain gravitation. These are the two types
of closed superstring, the two heterotic strings, and the open
superstring. Each of them is characterized by a single dimensionful
parameter, usually called $\alpha^\prime$, the inverse string
tension. This sets the scale for {\it all} observations and is as a
consequence not a measurable parameter --- it simply sets the scale for
units. Each string theory has a second dimensionless parameter, the
string coupling constant, that determines the strength with which
strings interact. This constant is freely specifiable, and thus
manifests itself as a massless scalar field in the theory, the
dilaton. However, like all other massless scalar fields, it must acquire
a mass through some quantum effects. The fact that such a massless
scalar field has not been observed argues in support of this
conclusion. Thus the string coupling constant must be determined
intrinsically by the theory --- in any given vacuum it is
calculable. So, all that is left is $\alpha^\prime$ which is
unobservable anyway. There still appears to be a choice of which string
theory one should pick. However, the discovery of M-theory shows that in
fact all string theories are equivalent and so no choice needs to be
made.

There are two issues associated today with the cosmological constant.
The first is to explain why the actual cosmological constant is tiny
compared to the amounts of vacuum energy generated by the electroweak
vacuum or the QCD vacuum or other sources of vacuum energy.  The second
issue is why there is apparently a residual non-zero small vacuum energy
which at the present epoch is of the same order as the contributions to
the total energy density.  The two issues are logically independent, and
could be physically independent.  We will not try to deal with the
second issue, even though it could be an anthropic question because of the
apparent coincidence that the cosmological energy density is of the same
order as the combined forms of matter at the present time even though
the two forms depend differently on time.  The first issue has been
discussed as an anthropic one~\cite{SW:cc}, \cite{SW:cc1}, \cite{SW:cc2}.
However, the string theory
point of view is outlined in the previous section.
Another possibility~\cite{GV:likely},~\cite{SW:r4} 
is that there might be very light scalars left over 
in the theory. Standard wisdom has it that such objects would by now have
been detected, but there is always the possibility that they couple 
so weakly to gravitation that they would not have been seen. 
What is interesting about the above  scenario is that the potential for such a
scalar field would look very much like an ordinary cosmological constant
except that the value of the scalar itself could very both in time and
space on cosmological distances or timescales. Under these circumstances,
it is thus a possibility that the effective cosmological constant here and now
is anthropically determined.

Another possibility at our present state of knowledge for understanding
how not to be uncomfortable with a universe that seems to be ``just so''
arises from the observation that universes could be arising with
different initial conditions and early histories, leading to different
parameters.  There are several approaches today to how universes might
begin~\cite{HH:wfu}~\cite{L:chi}~ \cite{Ve:sc} ~\cite{Vi:create}, and
of course additional approaches may be required.  Some or none of them
could be correct.  If many universes arise, various initial conditions
could lead to various resulting sets of parameters.  It is then like a
random lottery in that someone wins, and even though they may feel
singled out, from a broader viewpoint there is nothing special about
whomever won.

Finally, the vacuum structure of string theory is expected to be very
complicated. A
10D world is a consistent one as far as is known, and perhaps so are
many compactified ones.  There may be many stable solutions (local minima), 
each with
different dimensions, topological characteristics, and parameters.  Once
a universe falls into one of those minima the probability of tunneling
to a deeper minimum may be extremely small, so that the lifetime in many
minima is large compared to the lifetime needed for life to arise in
those minima. For a discussion of such phase transitions see Adams and
Laughlin~\cite{AL:paper}.
Then life would actually arise in those minima that were
approximately ``just so''.  Thus the ``just so'' issue is resolved by
having a large number of possible vacua in which universes can end up.
Eventually understanding of M-theory may reach the stage where it is
possible to calculate in practice all the possible vacua.  In each
vacuum, all of the quantities needed for a complete description of the
universe, including the masses and couplings that Hogan argues need to
be ``just so'', are calculable.

\section{Concluding Remarks}

We have argued that the usual anthropic arguments cannot be relevant
to understanding our world if string theory is the right approach to
understanding the law(s) of nature and the origins of the universe.
Our arguments are predicated on this hypothesis. If the type of unification
found in string theory is not an appropriate description of nature, then
we are back to the beginning in trying to understand why the universe has been
kind enough to us to allow us to live here.
If any parameters such as force strengths or quark masses
or the electron mass must be somehow adjustable and not fixed by the
theory in order to understand why the universe is ``just so'', then
string theory cannot be correct. We discuss various ways consistent
with string theory in which different universes with different
parameters could arise, so that the apparent ``just so'' nature of a
number of parameters can be understood.

\section{Acknowledgements}

We would like to thank the Institute for Theoretical Physics in Santa
Barbara, where this work was initiated, for its hospitality and support
in part by National science Foundation grant PHY94-07194,
and similarly the California Institute of Technology where it was completed.
GLK was supported in part by the US Department of Energy, and MJP 
partly supported by Trinity College Cambridge. We appreciate various
discussions with Fred Adams, Roger Blandford, Mike Duff, Craig Hogan
and Steven Weinberg.


\begin{thebibliography}{99}
\bibitem{BT:cap} J.D. Barrow and F.J. Tipler, \lq\lq The Anthropic
Cosmological Principle,'' Oxford University Press, 1986.
\bibitem{BC:anth1} B. Carter, \lq\lq Large Number Coincidences and the
Anthropic Principle in cosmology,'' in M.S. Longair, ``Confrontation
of Cosmological Theories with Observational Data,'' IAU, 1974.
\bibitem{D:ln} P.A.M. Dirac, ``A new basis for cosmology,'' Proc. Roy.
Soc. (London), {\bf A165}, 199, 1938.
\bibitem{RHD:ln} R.H. Dicke, ``Dirac's Cosmology and Mach's
Principle,'' Nature, {\bf 192}, 440, 1961.
\bibitem{CH:justso} C. Hogan, ``Why the Universe is Just So,''
astro-ph 9909295.
\bibitem{GLK:what} G.L. Kane, ``Supersymmetry'', Perseus Books,
Cambridge, MA, 2000.
\bibitem{PE:stable} P. Ehrenfest, Proc. Amst. Acad {\bf 20}, 200, 1917;
Ann. Phys. (Leipzig), {\bf 61}, 400, 1920.
\bibitem{SW:cc} S. Weinberg, ``The Cosmological Constant Problem,''
Rev. Mod. Phys. {\bf 61}, 1, 1989.
\bibitem{SW:cc1} S. Weinberg, in ``Theories of the Cosmological Constant,''
in ``Critical Dialogues in Cosmology,''  pp 195-203, ed N. Turok, 
World Scientific, Singapore, 1997.
\bibitem{SW:cc2} H. Martel, P. Shapiro and S. Weinberg, ``Likely Values of
the Cosmological Constant,'' Ap.J. {\bf 492}, 29-40, 1998.
\bibitem{GV:likely} J. Garriga and a. Vilenkin, ``On likely values of the
Cosmological Constant,'' astro-ph 9908115.
\bibitem{SW:r4} S. Weinberg, {\it in preparation}.
\bibitem{HH:wfu} J.B. Hartle and S.W. Hawking, Phys. Rev. D {\bf 28},
2960, 1983.
\bibitem{L:chi} A.D. Linde, ``Quantum Creation of an Inflationary Universe,''
Sov. Phys. JETP, {\bf 60}, 211, 1984.
\bibitem{Ve:sc} M. Gasperini and G. Veneziano, ``Towards a Nonsingular Pre-Big
Bang Cosmology,'' Nucl. Phys. {\bf B494}, 315, 1997.
\bibitem{Vi:create} A. Vilenkin, ``Quantum Creation of Universes,''
Phys. Rev. {\bf D30}, 509, 1984.
\bibitem{AL:paper} F.C. Adams and G. Laughlin, ``A dying Universe:
the long term fate and evolution of astrophysical objects.''
Rev. Mod. Phys. {\bf 69}, 337,
1997.
\bibitem{AL:book} F.C.Adams and G. Laughlin, ``Five Ages of the Universe:
Inside the Physics of Eternity,''
Freepress, NY 1999.
\bibitem{BG:elegant} B.R. Greene, ``The Elegant Universe: Superstrings, Hidden 
Dimensions, and the Quest for the Ultimate Theory.'' Norton, New York, 1999.
\bibitem{RC:anth} B.J. Carr and M.J. Rees, ``The anthropic principle
and the structure of the physical world,'' Nature, {\bf 278}, 605, 1979.
\bibitem{W:cy} P. Candelas, G.T. Horowitz, A.Strominger and E. Witten,
``Vacuum configurations for Superstrings,'' Nucl. Phys. {\bf B258}, 46, 1985.
\bibitem{BG:qg} B.R. Greene, ``Lectures on quantum geometry,'' Nucl. Phys.
Proc. Suppl. {\bf 41}, 92, 1995.
\bibitem{PC:roll} A.C. Avram, P. Candelas, D. Jancic and M. Mandelberg,
``On the Connectedness of Moduli Spaces of Calabi-Yau Manifolds,''
Nucl. Phys. {\bf B465}, 458, 1996.

\end{thebibliography}
\end{document}